\documentclass[%
reprint,
amsmath,amssymb,
aps,
prl,
]{revtex4-2}

\usepackage{graphicx}% Include figure files
\usepackage{bm}% bold math
\usepackage{siunitx}
\usepackage{physics}
\usepackage{xcolor}
\usepackage[colorlinks=true,linkcolor=blue,urlcolor=blue,citecolor=blue]{hyperref}

\newcommand{\Sr}{$^{87}$Sr}
\newcommand{\Er}{$E_r$}

\begin{document}

\preprint{}

%\title{Coherence time limit of an optical lattice clock}
% \title{Wannier-Stark Lattice Clock with 118 s Atomic Coherence and 1-s Atomic Instability of 1.5$\times$10$^{-18}$}
\title{Atomic Coherence of 2 minutes and Instability of 1.5\texorpdfstring{$\times 10^{-18}$}{E-18} at 1 s in a Wannier-Stark Lattice Clock}

\author{Kyungtae Kim, Alexander Aeppli, William Warfield, Anjun Chu, Ana Maria Rey, and Jun Ye}
\affiliation{$^1$~JILA, National Institute of Standards and Technology and the University of Colorado, Boulder, Colorado 80309-0440, USA \\
and Department of Physics, University of Colorado, Boulder, Colorado 80309-0390, USA}

\date{\today}
\begin{abstract}
We explore the limits of atomic coherence and measurement precision in a \Sr{} optical lattice clock. We perform a detailed characterization of key effects, including lattice Raman scattering and atomic collisions in a shallow lattice configuration, determining a 174(28) s $^3P_0$ clock state lifetime. Investigation of atomic coherence across a range of lattice depths and atomic densities reveals decoherence mechanisms related to photon scattering and atomic interaction. At a reduced density, we observe a coherence time of 118(9) s, approaching the fundamental limit set by spontaneous emission. Guided by this coherence understanding, we demonstrate a clock instability of 1.5$\times$10$^{-18}$ at 1 s in fractional frequency units. Our results are important for further advancing the state-of-the-art of an optical lattice clock for fundamental physics applications.
\end{abstract}
\maketitle

\textbf{\emph{Introduction}.}
Optical lattice clocks (OLCs) offer exceptional measurement precision by simultaneously interrogating a large number of atoms with a long coherence time~\cite{takamotoOpticalLatticeClock2005, boydOpticalAtomicCoherence2006, ludlowOpticalAtomicClocks2015}. The applications of OLCs range from timekeeping~\cite{dimarcqRoadmapRedefinitionSecond2024} to quantum sensing for fundamental physics~\cite{yeEssayQuantumSensing2024, bothwellResolvingGravitationalRedshift2022,zhengLabbasedTestGravitational2023,takamotoTestGeneralRelativity2020} and are a versatile platform for exploring many-body physics~\cite{ZhangSpectroscopic2014, bromleyDynamicsInteractingFermions2018, SonderhouseThermodynamics2020, hutsonObservationMillihertzlevelCooperative2024, milnerCoherentEvolutionSuperexchange2024}. Of fundamental importance in modern quantum science and technology is the scalability of a quantum system, and OLCs provide an ideal platform to explore relevant trade-offs for optimization. The use of many atoms reduces quantum projection noise (QPN) but inevitably introduces atomic interaction as a potential road block for both precision and accuracy. Spin squeezing provides a potential solution by providing better signal-to-noise with fewer atoms~\cite{robinsonDirectComparisonTwo2024}. Using an insulating quantum gas in a 3D optical lattice or optical tweezer arrays provides another route for number scaling~\cite{campbellFermidegenerateThreedimensionalOptical2017a,madjarovAtomicArrayOpticalClock2019,norciaSecondsscaleCoherenceOptical2019}. However, even minute interaction effects such as weak dipolar coupling~\cite{hutsonObservationMillihertzlevelCooperative2024} or superexchange spin interaction~\cite{milnerCoherentEvolutionSuperexchange2024} can noticeably affect clock operation. In a 1D Wannier-Stark lattice, we have engineered the interaction Hamiltonian to overcome a trade-off between systematics and atom number~\cite{aeppliHamiltonianEngineeringSpinorbit2022}. These efforts share a common goal: to realize atomic coherence time limited by the fundamental spontaneous emission while employing many atoms. 
% \red{Also tweezer experiments, atom interferometry~\cite{pandaCoherenceLimitsLattice2024}, MPQ optical qubits.} 

One major limitation to the observed coherence time in \Sr{} clock transition arises from Raman scattering of the lattice photons for individual atoms ~\cite{hutsonEngineeringQuantumStates2019,dorscherLatticeinducedPhotonScattering2018,youngHalfminutescaleAtomicCoherence2020}. The scattering events transfer populations from the $^3P_0$ state to the $^3P_1$ state, which then decay into the $^1S_0$ state. This process leads to a reduction in the contrast of Ramsey spectroscopy~\cite{niroulaQuantumSensingErasure2024}. It also populates the $^3P_2$ state, which has a large inelastic cross-section with $^3P_0$. The other source of decoherence is atomic interaction~\cite{campbellProbingInteractionsUltracold2009}. Although a large number of atoms $N$ is desired to reduce QPN, it degrades the coherence time through atomic interaction. 
% The collisional effect can be resolved by strongly confining atoms in a 3D optical lattice~\cite{campbellFermidegenerateThreedimensionalOptical2017a} to reach an interaction-dominated insulator regime, but at the cost of increased lattice scattering rate in order to suppress tunneling~\cite{hutsonEngineeringQuantumStates2019,milnerCoherentEvolutionSuperexchange2024}. 
A large beam waist, gravity-induced Wannier-Stark 1D optical lattice~\cite{lemondeOpticalLatticeClock2005} allows us to operate the clock at a lattice depth of only a few photon recoil energy $E_r$, which greatly reduces the lattice photon scattering as well as atomic density. 

Previously, we investigated how the spin-orbit coupling~\cite{kolkowitzSpinOrbitCoupled2016,bromleyDynamicsInteractingFermions2018} in a Wannier-Stark OLC introduces off-site $s$-wave interaction~\cite{aeppliHamiltonianEngineeringSpinorbit2022}. Near a specific optical lattice depth $U_0\sim$10$E_r$, we null the mean interaction strength, enabling us to utilize a large $N$ without losing metrological precision. This is essential for resolving sub-mm gravitational redshift~\cite{bothwellResolvingGravitationalRedshift2022} and reducing systematic uncertainties~\cite{kimEvaluationLatticeLight2023,aeppliClock10192024}. A natural next step is to explore how these interactions affect the coherence time.

In this work, we study the effect of lattice light scattering and atomic collisions on the clock performance for different lattice depths and demonstrate an atomic coherence time of $\sim$2 minutes. Raman scattering leads to the accumulation of the population in different nuclear spin states of $^1S_0$. Resulting `spectator' atoms collide via strong $s$-wave interactions with those in clock states, dominating the decoherence rate. With in-situ imaging ~\cite{martiImagingOpticalFrequencies2018} of the atomic distribution, we infer the coherence time extrapolating to a zero density limit. And when this limit is further extrapolated to $U_0$=0, we find that the atomic decoherence is in agreement with the limit set by the natural lifetime of $^3P_0$ and the black body radiation (BBR) from the environment. Furthermore, we use this system to investigate the intrinsic clock precision and demonstrate instability of $1.5\times10^{-18}$ at 1 s. 

Detailed description of the experimental apparatus can be found in Refs.~\cite{aeppliClock10192024,aeppliHamiltonianEngineeringSpinorbit2022,bothwellResolvingGravitationalRedshift2022}. We prepare the atoms in $\ket{^1 S_0 \equiv g, m_F=-5/2}$ at $U_0$=20\Er{}. For the population decay measurement, we use a $\pi$-pulse to populate $\ket{^3 P_0 \equiv e, m_F=-3/2}$ and remove the remaining population in $\ket{g}$ using a strong $^1S_0\leftrightarrow{}^1P_1$ transition at 461~\unit{\nano\meter}. Then, we adiabatically ramp the lattice to the desired $U_0$ in 50~\unit{\milli\second} and hold it with a variable time and measure the atomic population. To measure the coherence time of the clock transition, $\ket{g, m_F=-5/2}\leftrightarrow\ket{e, m_F=-3/2}$, we observe the contrast of the Ramsey fringe with the varying dark time. Finally, we use the imaging spectroscopy method~\cite{martiImagingOpticalFrequencies2018} to estimate the frequency measurement noise contributed by the atoms.

\textbf{\emph{Population decay}}. Environmental perturbations, such as lattice photon scattering, can extract information from an atom~\cite{uysDecoherenceDueElastic2010,ozeriHyperfineCoherencePresence2005}, and consequently, any state-dependent perturbation can cause decoherence of the clock superposition. For example, half of the $e \rightarrow g$ decay rate directly contributes to the decoherence rate. The use of a magic wavelength in OLCs protects coherence, by removing the information carried out by the photon~\cite{hutsonEngineeringQuantumStates2019,dorscherLatticeinducedPhotonScattering2018}, as well as minimizing the effect from atomic motion. State-independent trap loss does not directly affect the coherence time, but it can have an indirect impact by requiring an increase of the initial $N$ to achieve a reasonable signal-to-noise at the detection stage.

The population dynamics of the atoms in the optical lattice can be described by the following rate equation.
\begin{equation} \label{eqn:pop_decay}
    \begin{aligned}
    \dot{N_e} &= -\Gamma_e N_e - \Gamma_L N_e - \tilde{\Gamma}_{ee} \kappa N_e^2, \\
    \dot{N_g} &= -\Gamma_g N_g + \Gamma_L N_e,
    \end{aligned}
\end{equation}
where $N_{e(g)}$ is the atom number in $\ket{e(g)}$, $\Gamma_{e(g)}$ is the single-body loss rate for $e(g)$, $\tilde{\Gamma}_{ee}$ is the two-body loss rate of $\ket{e}$, and $\kappa$ is a prefactor that converts $N_e$ to density for a two-body loss process~\cite{bishofInelasticCollisionsDensitydependent2011}. $\Gamma_L(U) = \Gamma_L(0) + (\partial_U\Gamma_L)U$ is the rate for $e \rightarrow g$, where $U$ is the averaged lattice depth. To take into account both the axial and radial spread of the atoms~\cite{ushijimaOperationalMagicIntensity2018,kimEvaluationLatticeLight2023}, we use $U=\eta(1)U_0 - \eta(1/2)\sqrt{U_0}$ and $\eta(j) = (1+jk_BT_r/U_0E_r)^{-1}$, where $k_B$ is the Boltzmann constant, $T_r$ is the radial temperature, and $U_0$ is the peak lattice depth. We measure the dynamics starting in  $\ket{e}$ and fit the data to equation~\eqref{eqn:pop_decay} using a least squares method to extract the decay rates. 

\begin{figure}[th!]
    \includegraphics[width=8.5cm]{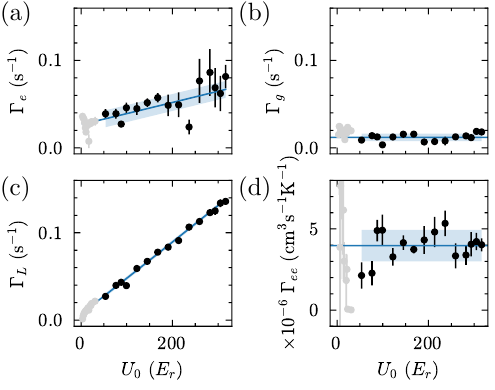}
    \caption{\label{fig:lattice_raman}
    Lattice depth dependent population decay rates. (a) Single-body loss rate of the excited state. (b) Single-body loss rate of the ground state. (c) $e \rightarrow g$ pumping rate. (d) Two-body loss rate of the excited state. Each horizontal axis represents the peak lattice depth, $U_0$. We use the lattice depth greater than the 50\Er{} (black markers) for fitting the data, in order to avoid complications of the model such as Raman scattering induced loss and lattice intensity noise. The extracted coefficients are summarized in Tab.~\ref{tab:pop_decay}. The error bars shows the 68\% confidence interval. The blue lines are fitted curves and the shades are their uncertainties.
    % Fitted coefficients
    % gamma_e_bg Fitted parameters slope: 1.32e-04 +- 3.12e-05, intercept: 2.70e-02 +- 4.01e-03
    % gamma_g_bg Weighted average: 1.20e-02 +/- 4.34e-03
    % gamma_L Fitted parameters slope: 4.30e-04 +/- 7.01e-06, intercept: 8.12e-03 +/- 8.22e-04
    % gamma_ee Weighted average: 3.97e-06 +/- 9.69e-07
    }
\end{figure}

\begin{table}[]
    \centering
    \renewcommand{\arraystretch}{1.3}
    \begin{tabular}{c c}
    \hline \hline
    Quantity & Value\\
    \hline
        $\Gamma_e (U)$ & $(1.3(3)\times 10^{-4} U/E_r + 2.7(4) \times 10^{-2} )$ \unit{s^{-1}} \\
        $\Gamma_g$ & $1.2(4)\times 10^{-2}$ \unit{s^{-1}} \\
        $\Gamma_L (U)$ & $(4.30(7) \times 10^{-4} U/E_r + 8.1(8) \times 10^{-3})$ \unit{s^{-1}}\\
        $\tilde{\Gamma}_{ee}$ & $4(1) \times 10^{-6}$ \unit{cm^{-3}s^{-1}K^{-1}} \\
        $1/\Gamma_{nat}$ & 174(28)~\unit{s} \\
    \hline
    \end{tabular}
    \caption{\label{tab:pop_decay} Summary of the population decay rate measurement. $U$ is average lattice depth in units of \Er{}$=(h/\lambda_L)^2/2M$, where $\hbar$ is Planck constant, $\lambda_L$ is the lattice wavelength, and $M$ is the mass of \Sr{}.}
\end{table}

Figure~\ref{fig:lattice_raman} and Tab.~\ref{tab:pop_decay} present the measurement results. Raman scattering drives population into the $^3P_1$ and $^3P_2$ states~\cite{dorscherLatticeinducedPhotonScattering2018} and then the $^3P_1$ state quickly decays to the $^1S_0$ state with a rate $\Gamma_L$. The value of $\Gamma_L(0) = 8.1(8)\times 10^{-3}$~\unit{s} represents the limit set by the combined effect of the spontaneous decay of $^3P_0$ and the BBR scattering rate. After subtracting the contribution of the BBR of $2.36 \times 10^{-3}$~\unit{s^{-1}}~\cite{yasudaLifetimeMeasurementMetastable2004,dorscherLatticeinducedPhotonScattering2018, aeppliClock10192024}, we obtain a lifetime of the $^3P_0$ state of $1/\Gamma_{nat}$ = 174(28)~\unit{s}. This is in agreement with previous measurements~\cite{boydNuclearSpinEffects2007, dorscherLatticeinducedPhotonScattering2018, luDeterminingLifetime52024}, and longer than \cite{munizCavityQEDMeasurementsSr2021}. $\Gamma_g$ shows no dependence on $U_0$, suggesting the background gas collision as the dominant $\ket{g}$ loss mechanism. On the other hand, $\Gamma_e$ shows a linear dependence on $U_0$. We attribute this dependence to the Raman scattering into $^3 P_2$, which has a large inelastic cross-section with the clock states. The value of $\tilde{\Gamma}_{ee}$ is consistent with a previous measurement~\cite{bishofInelasticCollisionsDensitydependent2011} and larger than the value reported in Ref.~\cite{lisdatCollisionalLossesDecoherence2009}. 
% BBR scattering rate at 22C: 424.64 ± 0.98 s or 2.36 × 10−3 s−1

\begin{figure*}[th!]
    \includegraphics[]{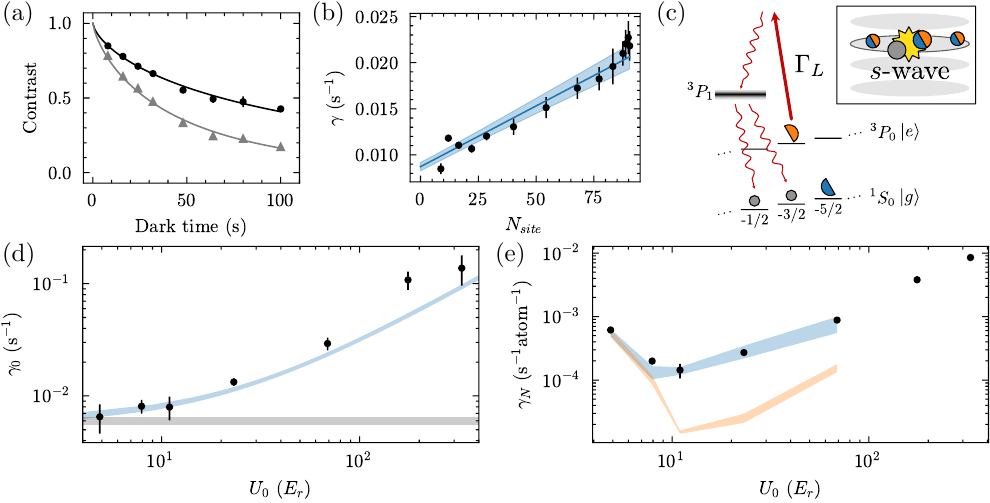}
    \caption{\label{fig:dens_dep} 
    Collisional interactions and the atomic coherence time. (a) Contrast of the Ramsey fringe as a function of the dark time for two different mean atom number per site cases. The circle (triangle) is for $N_{site}=9(90)$. Solid lines are a fit to a stretched exponential model. The coherence time, $\gamma^{-1}$ for $N_{site}=9$ is 118(9) s at 11\Er{}. (b) Density dependence of the contrast decay rate. The blue solid line is a fit to a linear curve. (c) A cartoon illustrates the generation of spectator atoms (gray) via the lattice Raman scattering. The semicircles represent halves of a superposition. The spectator atoms lead to additional phase diffusion through the $s$-wave collision. (d) Lattice depth dependence of $\gamma_0$. The bands are estimation of the contrast decay time. The gray band is $(\Gamma_{nat} + \Gamma_{BBR})/2$. The blue band is $(\Gamma_R U + \Gamma_{nat} + \Gamma_{BBR})/2$. (e) Lattice depth dependence of $\gamma_N$. The band shows DDTWA simulations. The blue band with spectator atoms and the orange band without interaction effect from the spectator atoms. The error bars show the 68\% confidence interval. 
    % \red{I put the orange band back. Without it, it is hard to make the story for the spectator atoms convincing.}
    % Nonlinearity treated with fitting upper and lower 2/3 of the N_site to another linear curves and take the differnece in the fitting parameter, added them in quadratrue. 
    % Also need to describe binning size changing procedure for more robust extrapolation to zero density limit.
    }
    %Coherence time for N_tot=9: 117.74 s (8.7 s)
    %Coherence time for N_tot=90: 45.83 s (3.6 s)
\end{figure*}

\textbf{\emph{Coherence time}}. We investigate the coherence time of the atomic ensembles using Ramsey interferometry. Because the coherence time of the atoms exceeds that of laser~\cite{oelkerDemonstration4810172019, matei15LasersSub102017}, the atom-laser phase is randomized at the readout. As a result, we repeat the experiment multiple times for a given dark time, $T_{dark}$, and measure the change in peak-to-peak excitation fraction as an estimate of the contrast, $C$. To avoid bias, we use only the sub-ensemble regions where the QPN is limited to a maximum of $\sim5\%$ excitation fraction. We model the contrast decay trajectory (Fig.~\ref{fig:dens_dep}(a)) with an empirical stretched exponential, $C(T_{dark}) = C(0)\exp[-(\gamma T_{dark})^\alpha]$, where $\{C(0),~\gamma,~\alpha\}$ are the fit parameters, and $\gamma$ represents the contrast decay rate. The uncertainty is estimated via bootstrapping. The coherence time for with an atom number per lattice site $N_{site}=9$ is 118(18) s for $U_0$ = 11\Er{}, plotted as a black curve in Fig.~\ref{fig:dens_dep}(a).

\begin{figure}[th!]
    \includegraphics[width=8.5cm]{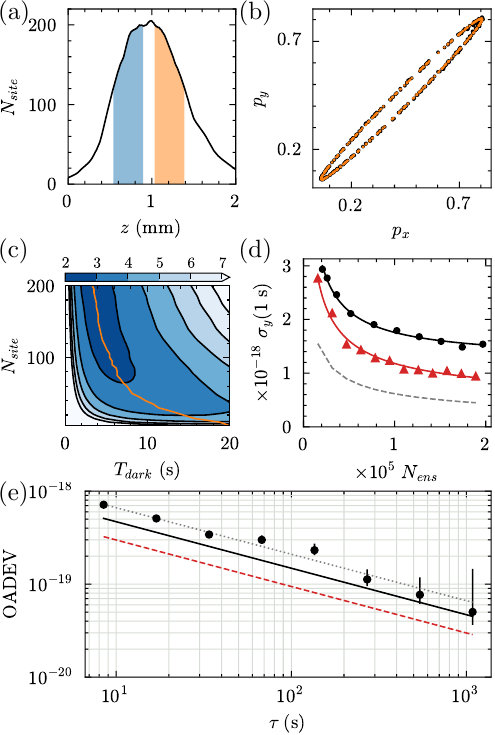}
    \caption{\label{fig:stability}
    Estimation of atomic contribution to the clock stability. 
    (a) 1D image of the atomic cloud. $z$ is the coordinate along the gravity. We estimate the frequency difference between two regions (e.g., the blue and the orange) using ellipse fitting.
    (b) Parametric plot of the excitation fractions of two regions. The black markers are the experimental data and the orange dots are fitted ellipse. 
    (c) A single lattice site's $\sigma_{y}(1~\unit{s})$ in units of $10^{-17}$ as a function of $N_{site}$ and $T_{dark}$ from DDTWA. The orange line indicates the optimal $T_{dark}$ as a function of $N_{site}$. (d) 1-\unit{s} instability for different $N_{ens}$. The black circles are experimental data and the red triangles are theoretical predictions. The solids lines are heuristic fits. The gray dashed line shows the theoretical prediction from Eqn.~\eqref{eqn:phi_to_y} for coherent spin states.
    (e) Overlapping Allan deviation for $\sigma_{y}(\tau)$. The black markers are comparison instability $\sigma_{y}^{rel}$ and the gray line the fit. The black solid line is the single clock instability $\sigma_{y}(\tau) = 1.5\times10^{-18}/\sqrt{\tau/\unit{s}}$. The red dashed line is theoretical prediction. The error bars shows the 68\% confidence interval. 
    } 
\end{figure}

The extracted $\gamma$ shows a strong density dependence (Fig.~\ref{fig:dens_dep}(b)). We fit the data to a linear curve, $\gamma = \gamma_0 + \gamma_N N_{site}$. Here, $\gamma_0$ represents the contrast decay rate at the single-atom regime, and $\gamma_N$ quantifies the collisional interaction effect. We subdivide the image of the atomic distribution for different values of $N_{site}$ (see also Fig.~\ref{fig:stability}(a)).

Figures~\ref{fig:dens_dep}(d, e) summarize the dependence of $\gamma_0$ and $\gamma_N$ on $U_0$. $\gamma_0$ is dominated by the lattice Raman scattering rate $\Gamma_R$, the single photon scattering rate from BBR $\Gamma_{BBR}\sim  1/164$ $ \unit{s^{-1}}$, and the natural lifetime of the excited state $\Gamma_{nat}$. We use the result from the previous section $\Gamma_R=5.6(3)\times10^{-4}$~\unit{s^{-1}}\Er{}$^{-1}$. As shown in Fig.~\ref{fig:dens_dep}(d), $\gamma_0$ is mainly limited by $\Gamma_R$ at high $U_0$, and converges to a value close to the sum of $\Gamma_{nat}$ and $\Gamma_{BBR}$ as $U_0$ approaches 0. We find that the observed $\gamma_0$ is captured by a simple estimation of $(\Gamma_R U + \Gamma_{nat} + \Gamma_{BBR})/2$. % BBR single photon scattering rate: 163.57 ± 0.38 s

In contrast to $\gamma_0$, $\gamma_N$ shows a non-monotonic dependence on $U_0$ (Fig.~\ref{fig:dens_dep}(e)). At shallow depths, delocalization between adjacent lattice sites introduces off-site $s$-wave interactions via spin-orbit coupling~\cite{aeppliHamiltonianEngineeringSpinorbit2022}, which dominates the decoherence. As the lattice depth increases, the $s$-wave channel is suppressed, and the on-site $p$-wave contribution grows with density. However, the latter effect does not explain the data quantitatively with the limited strength of $p$-wave interaction (the orange band).

Lattice Raman scattering introduces additional decoherence through the generation of spectator atoms~(Fig.~\ref{fig:dens_dep} (c)). The photon scattering events populate various nuclear spin states in $g$ that are distinguishable from the clock state. These spectator atoms interact with the clock atoms via strong on-site $s$-wave collisions, becoming a dominant source of decoherence. In addition, the stochastic generation of the spectator atoms introduces further fluctuations in the clock phase~\cite{liFundamentalLimitPhase2020}. These mechanism are supported by theoretical simulations based on a dissipative discrete truncated Wigner approximation (DDTWA)~\cite{SeeSupplementalMaterial}. In Fig.~\ref{fig:dens_dep}(e), we show the simulation result of $\gamma_N$ with and without the spectator atoms. For deeper lattices, the decoherence induced by the presence  of spectator atoms becomes prominent, which limits the use of large atom numbers. We note that the simulation shows a nonlinear dependence of $\gamma$ on $N_{site}$~\cite{SeeSupplementalMaterial}. To account for small nonlinearity, we fit the line for two different ranges, $[0, (2/3)\operatorname{max}(N_{site})]$ and $[(1/3)\operatorname{max}(N_{site}), \operatorname{max}(N_{site})]$, and take the difference as extra uncertainties for $\gamma_0$ and $\gamma_N$. The same treatment is applied to the theoretical simulation when extracting $\gamma_N$ and its range is indicated by the bands.
We exclude theoretical simulations for $U_0>10^2E_r$ due to extra sources of decoherence in this regime such as atoms in higher bands not captured by our model. 
% To compare with experimental data, we find the range of the theoretically estimated $\gamma_N$ as bands~\cite{SeeSupplementalMaterial}.

%  We parameterize the interaction strength between the clock states and the spectator as $\chi_{spec}$ to fit the data. This factor effectively describes the initial atomic polarization impurity. In Fig.~\ref{fig:dens_dep}(d), we see that $\gamma_N$ is an order of magnitude different at high lattie depths if we set $\chi_{spec}=0$, shows that the dominant decoherence source is coming from the interaction with the spectator atoms.

\textbf{\emph{Imaging spectroscopy}}.
To estimate the atomic contribution for the clock instability beyond the laser coherence time, we perform a synchronous clock comparison by using a Ramsey protocol achieved through imaging spectroscopy~\cite{martiImagingOpticalFrequencies2018}. The frequency difference of the two regions~(Fig.~\ref{fig:stability}(a)) is reflected as a correlation between the excitation fractions, resulting in a parametric plot with the shape of an ellipse~(Fig.~\ref{fig:stability}(b)). The opening angle of the ellipse,  $\phi$, related to the frequency difference of the two regions is obtain from the ellipse fit. The QPN contribution to the variance of $\phi$ can be estimated as~\cite{martiImagingOpticalFrequencies2018}, 
\begin{equation} \label{eqn:instability}
    \operatorname{var} (\phi) = 
    \frac{4}{C^2}\left(\int_0^{2 \pi} \frac{d \theta}{2 \pi} \frac{1}{\sum_{i=x,y}\csc ^2 (\theta_i) \operatorname{var} (p_i)}\right)^{-1}.
\end{equation}
Here, $p_{x, y}=(1+C\cos(\theta_{x,y}))/2$ is the excitation fractions for each region with $\theta_{x,y}=\theta\mp\phi/2$, $C$ is the contrast, $\theta$ is the phase of the laser which assumes to be uniformly distributed, and $\phi$ is the Ramsey phase difference between two regions~(Fig.~\ref{fig:stability}(a)). For coherent spin states, the variance $\operatorname{var} (p_i) = p_i(1-p_i)/N_{ens}$, where $N_{ens}$ is the number of atoms in one ensemble. Note that Eqn.~\eqref{eqn:instability} is a good approximation for the classical Cram\'er-Rao bound for large atom numbers and not so small $\phi$~\cite{pezzeQuantumMetrologyNonclassical2018, SeeSupplementalMaterial}. The QPN contribution to the clock instability can be, therefore, estimated as 
\begin{equation} \label{eqn:phi_to_y}
\sigma_{y} (\tau) = \frac{\sigma_{y}^{rel} (\tau)}{\sqrt{2}} = \frac{\sqrt{ \operatorname{var} (\phi)} }{ 2 \pi \nu_0 T_{dark} \sqrt{2\tau/T_{cycle}}},
\end{equation}
where $\sigma_{y}^{rel}$ is relative (comparison) instability between two regions, $\nu_0$ is the frequency of the clock transition, $T_{cycle}$ is the experimental cycle period, and $\tau$ is the averaging time. The reduction by a factor of $\sqrt{2}$ accounts for the independent contribution from the two regions.

The contrast decay limits the achievable sensitivity with increasing $T_{dark}$ and $N_{site}$. This competition results in a minimum instability at specific $T_{dark}$ for a given $T_{cycle}$ and $N_{site}$. Figure~\ref{fig:stability}(c) presents such a parametric contour plot of $\sigma_{y} (\tau)$ based on DDTWA for $U_0=11E_r$, lattice depth at which we see a minimal density-dependent contrast decay, and hence the best stability. The plot assumes a magnetic field gradient of 12.7~\unit{mHz/mm}, a separation between adjacent lattice sites  of 260~\unit{\micro\meter} and a phase accumulation $\phi$ linear with $T_{dark}$. The experimental dead time is accounted as $T_{cycle} = T_{dark} + 1.5~\unit{s}$. The density profile of the sample (Fig.~\ref{fig:stability}(a)) suggests using $T_{dark}=4\sim8~\unit{s}$.

In Fig.~\ref{fig:stability}(e), we present the Allan deviation for the comparison and a single clock instability, under $T_{dark}=7~\unit{s}$ and $T_{cycle}=8.48~\unit{s}$ with 313 realizations. A jackknifing method is used to generate series of $\phi$ and compute the Allan deviation~\cite{martiImagingOpticalFrequencies2018}. Subsequently, we convert $\phi$ to $\sigma_{y}^{rel}$. The fit to data, $\propto 1/\sqrt{\tau}$ (the gray dotted line), along with a single clock $\sigma_y (1~\unit{s})=1.5\times10^{-18}$ (the black solid line) are plotted with the theoretical QPN contribution for $\sigma_y (1~\unit{s})$ of $9.4\times 10^{-19}$ (the red dashed line).

The observed instability is 50\% larger than the theoretical estimate. To quantify the difference, we vary the bin size of the image (Fig.~\ref{fig:stability}(a)) to change the atom number per ensemble, $N_{ens}$, and estimate the instability for each $N_{ens}$ (Fig.~\ref{fig:stability}(d)). We observe that the experimental value starts to saturate after $N_{ens} \approx 5\times10^4$. We fit a heuristic curve $\sigma_y(1~\unit{s})=\sqrt{a^2/N_{ens} + b^2}$ to the data, where $a$ and $b$ are the fit parameters (solid lines). For the experimental data (black circles), $a=5.6\times10^{-16},~b= 1.8\times10^{-18}$ and for the theoretical prediction (red triangles ), $a=4.8\times10^{-16},~b= 6.5\times10^{-19}$. This suggests that the observed instability is limited by an atom number-independent noise source, to be investigated in the future. We note that ellipse fitting can introduce additional noise and bias depending on the method~\cite{corgierSqueezingenhancedAccurateDifferential2025, stocktonBayesianEstimationDifferential2007,ecknerRealizingSpinSqueezing2023}. We test the fitting method using simulated data; see~\cite{SeeSupplementalMaterial} for more details. We also emphasize that the theoretically simulated $\sigma_y$ (red) is larger than that predicted from coherent spin states (gray), indicating excess noise from the spectator atoms.

% popt_theory: [-4.85019645e-16  6.58532418e-19]
% popt experiment: [ 5.56676214e-16 -1.75706572e-18]

% We employ two different fitting methods: least squares with geometric distance (LSGD) and a Bayesian analysis. To determine the uncertainty of each of the methods we use  the jackknife method~\cite{martiImagingOpticalFrequencies2018} for the former and both jackknifing and the standard deviation of the posterior distribution for the latter. The Allan deviation results in  Fig.~\ref{fig:stability}(d, e) used the LSGD for both experimental data and theoretical calculations.  See \cite{SeeSupplementalMaterial} for more details.

\textbf{\emph{Conclusion.}}
We report a \Sr{} OLC coherence time of about 2 minutes in a shallow depth, low density sample. We find that the decoherence is dominated by the combination of lattice Raman scattering and atomic collisions. Furthermore, we demonstrate a single atomic region instability of $1.5\times10^{-18}/\sqrt{\tau/\unit{s}}$. Our findings contribute to a better understanding of the stability limits of state  of the art OLCs  and pave the way for future advancements in its development for fundamental physics applications~\cite{chuExploringDynamicalInterplay2025, yeEssayQuantumSensing2024, kolkowitzGravitationalWaveDetection2016, zychQuantumInterferometricVisibility2011}.

% \section{Acknowledgment}
\textbf{\emph{Acknowledgment.}}
We thank A. Cao, A. Kaufman, and Y. Yang for useful discussions and review of this manuscript, and M. Miklos for discussion about ellipse fitting. Funding support is provided by NSF QLCI OMA-2016244,  DOE National Quantum Information Science Research Centers - Quantum Systems Accelerator, V. Bush Fellowship, Physics Frontier Center PHY-2317149, AFOSR  FA9550-24-1-0179 and NIST. 

\bibliography{main}
\end{document}

%%%%%%%%%%  Custom commands %%%%%%%%%%
\newcommand{\Sr}{${}^{87}$Sr}
\newcommand{\Er}{$E_{r}$}
\renewcommand{\thefigure}{S\arabic{figure}}\makeatother
\renewcommand{\thetable}{S\arabic{table}}\makeatother
\newcommand{\red}[1]{\textcolor{red}{#1}}

%%%%%%%%%%%%%%%%%%%%%%%%%%%%%%%%%%%%%%

% \preprint{}
\title{Supplementary Material: Atomic Coherence of 2 minutes and Instability of 1.5$\times$10$^{-18}$ at 1 s in a Wannier-Stark Lattice Clock}% Force line breaks with \\
\author{Kyungtae Kim, Alexander Aeppli, William Warfield, Anjun Chu, Ana Maria Rey, and Jun Ye}
\affiliation{$^1$~JILA, National Institute of Standards and Technology and the University of Colorado, Boulder, Colorado 80309-0440, USA \\
and Department of Physics, University of Colorado, Boulder, Colorado 80309-0390, USA}

\maketitle

\section{DDTWA calculation for Wannier-Stark lattice clock}
In the main text, we perform dissipative discrete truncated Wigner approximation (DDTWA) in the calculation. Our method combines DTWA \cite{Schachenmayer2015} with quantum jumps, which is similar to Ref.~\cite{Singh2022} and is generalized beyond two-level atoms. An alternative method using stochastic differential equations is discussed in Ref.~\cite{Huber2022}.

Firstly, we briefly explain the quantum jump method. We consider the Lindblad master equation,
\begin{equation}
    \frac{d}{dt}\hat{\rho} =  -\frac{i}{\hbar} [\hat{H},\hat{\rho}] + \sum_{j=1}^{n_{\rm tot}}\bigg(\hat{L}_j\hat{\rho}\hat{L}_j^{\dag}-\frac{1}{2}(\hat{L}_j^{\dag}\hat{L}_j\hat{\rho}+\hat{\rho}\hat{L}_j^{\dag}\hat{L}_j)\bigg),
\end{equation}
where $\hat{H}$ is the Hamiltonian, and $\hat{L}_j$ are the quantum jump operators. We define the non-Hermitian Hamiltonian 
\begin{equation}
    \hat{H}_{\rm NH} = \hat{H}-\frac{i\hbar}{2}\sum_{j=1}^{n_{\rm tot}} \hat{L}_j^{\dag}\hat{L}_j,
\end{equation}
the quantum jump method for solving Lindblad master equation can be described in the following way:
\begin{itemize}
    \item Choose a random number $r_1$ based on the uniform distribution between $0$ and $1$. Starting from a pure state $|\psi(0)\rangle$, evolve the wave function under the non-Hermitian Hamiltonian $\hat{H}_{\rm NH}$, 
    \begin{equation}
        |\psi^{(1)}(\tau)\rangle = \exp\bigg(-i\frac{\hat{H}_{\rm NH}\tau}{\hbar}\bigg)|\psi(0)\rangle,
    \end{equation}
    a quantum jump occurs at time $\tau$ such that $\langle \psi^{(1)}(\tau)|\psi^{(1)}(\tau)\rangle = r_1$.

    \item Choose a random number $r_2$ based on the uniform distribution between $0$ and $1$. We choose the jump operator $\hat{L}_n$ if $n$ is the smallest integer satisfying
    \begin{equation}
        \sum_{j=1}^n P_{j}(\tau) \geq r_2, \quad P_j(\tau) = \frac{\langle \psi^{(1)}(\tau)|\hat{L}_j^{\dag}\hat{L}_j|\psi^{(1)}(\tau)\rangle}{\sum_{j=1}^{n_{\rm tot}} \langle \psi^{(1)}(\tau)|\hat{L}_j^{\dag}\hat{L}_j|\psi^{(1)}(\tau)\rangle},
    \end{equation}
    and the wave function becomes
    \begin{equation}
        |\psi (\tau)\rangle = \frac{\hat{L}_n|\psi^{(1)}(\tau)\rangle}{\sqrt{\langle \psi^{(1)}(\tau)|\hat{L}_n^{\dag}\hat{L}_n|\psi^{(1)}(\tau)\rangle}}.
    \end{equation}

    \item Repeat the procedure above with initial state $|\psi(\tau)\rangle$ until the final simulation time is reached.
\end{itemize}

To combine with DTWA, we modify the procedure above in the following way:
\begin{itemize}
    \item Replace the wave function evolution under non-Hermitian Hamiltonian $\hat{H}_{\rm NH}$ by DTWA evolution for a set of single-atom operators $\hat{O}_k$ and the identity operator $\hat{I}$. This leads to mean-field equations for expectation values $\langle \psi^{(1)}|\hat{O}_k|\psi^{(1)}\rangle$ and the wave function norm $\langle \psi^{(1)}|\psi^{(1)}\rangle$, with random sampling of initial conditions \cite{Schachenmayer2015,zhu2019generalized}.
    The random sampling of initial conditions is based on the following rules: Suppose $\hat{O}_k$ has eigenvalues $\alpha_j^{(k)}$ with corresponding eigenvectors $|\alpha_j^{(k)}\rangle$, we set $\langle \psi^{(1)}(0)|\hat{O}_k|\psi^{(1)}(0)\rangle = \alpha_j^{(k)}$ with probability $P_j^{(k)}=|\langle\psi^{(1)}(0)|\alpha_j^{(k)}\rangle |^2$. Note that we always have $\langle \psi^{(1)}(0)|\psi^{(1)}(0)\rangle=1$.
    \item When a quantum jump occurs, we reset the initial conditions based on the selected quantum jump operator and then perform DTWA evolution. We will repeat the procedure until final simulation time is reached.
\end{itemize}

Now we would like to apply the method described above to numerically simulate the dynamics in the Wannier-Stark lattice clock. Following the discussions in Ref.~\cite{aeppliHamiltonianEngineeringSpinorbit2022}, we have
\begin{equation}
    \hat{H} = \hat{H}_{\rm twolevel} + \hat{H}_{\rm spectator},
\end{equation}
where 
\begin{equation}
    \hat{H}_{\rm twolevel}=\sum_{\substack{\mathbf{n}\mathbf{m}\\(\mathbf{n}\neq\mathbf{m})}}\bigg[J_{\mathbf{n}\mathbf{m}}\hat{\mathbf{S}}_{\mathbf{n}}\cdot\hat{\mathbf{S}}_{\mathbf{m}}+\chi_{\mathbf{n}\mathbf{m}}\hat{S}^z_{\mathbf{n}}\hat{S}^z_{\mathbf{m}}+D_{\mathbf{n}\mathbf{m}}(\hat{S}^x_{\mathbf{n}}\hat{S}^{y}_{\mathbf{m}}-\hat{S}^y_{\mathbf{n}}\hat{S}^{x}_{\mathbf{m}})+\frac{C_{\mathbf{n}\mathbf{m}}}{2}(\hat{S}^z_{\mathbf{n}}\hat{N}^{eg}_{\mathbf{m}}+\hat{N}^{eg}_{\mathbf{n}}\hat{S}^z_{\mathbf{m}})\bigg],
\end{equation}
\begin{equation}
    \hat{H}_{\rm spectator}=\sum_{\substack{\mathbf{n}\mathbf{m}\\(\mathbf{n}\neq\mathbf{m})}}\bigg[\frac{K^a_{\mathbf{n}\mathbf{m}}}{2}(\hat{S}^z_{\mathbf{n}}\hat{N}^{a}_{\mathbf{m}}+\hat{N}^{a}_{\mathbf{n}}\hat{S}^z_{\mathbf{m}})+\frac{K^b_{\mathbf{n}\mathbf{m}}}{2}(\hat{S}^z_{\mathbf{n}}\hat{N}^{b}_{\mathbf{m}}+\hat{N}^{b}_{\mathbf{n}}\hat{S}^z_{\mathbf{m}})\bigg].
\end{equation}
Here, $\mathbf{n}=\{n_x,n_y,n\}$, with $n_x, n_y$ the number of radial harmonic oscillator modes and $n$ the index of lattice sites where the Wannier-Stark state centered at. 
$\hat{H}_{\rm twolevel}$ is acting between clock states $|e\rangle\equiv |{}^3P_0,m_F=-3/2\rangle$ and $|g\rangle\equiv|{}^1S_0,m_F=-5/2\rangle$, so we can defined $\hat{S}_{\mathbf{n}}^{x,y,z}$ are spin-$1/2$ operators for an occupied mode $\mathbf{n}$, and $\hat{N}^{eg}_{\mathbf{n}}$ describes the sum of population in $|e\rangle$ and $|g\rangle$ states for mode $\mathbf{n}$. 
We only include occupied modes in the summation, and they are sampled from thermal distribution based on the cloud temperature.
For simplicity, we drop all the terms in the Hamiltonian acting identically for $|g\rangle$ and $|e\rangle$ states since they have no contributions to the clock coherence and stability.

The Hamiltonian parameters in $\hat{H}_{\rm twolevel}$ has already been discussed in Ref.~\cite{aeppliHamiltonianEngineeringSpinorbit2022}, which includes contributions from both on-site $p$-wave interactions and nearest-neighbor $s$-wave interactions. For completeness, here we list them as follows:
\begin{equation}
    \begin{gathered}
    J_{\mathbf{n}\mathbf{m}} = (V^{eg}_{\mathbf{n}\mathbf{m}}-U^{eg}_{\mathbf{n}\mathbf{m}})\cos[(n-m)\varphi], \\ \chi_{\mathbf{n}\mathbf{m}} = (V^{ee}_{\mathbf{n}\mathbf{m}}+V^{gg}_{\mathbf{n}\mathbf{m}}-2V^{eg}_{\mathbf{n}\mathbf{m}}) + (V^{eg}_{\mathbf{n}\mathbf{m}}-U^{eg}_{\mathbf{n}\mathbf{m}})\Big(1-\cos[(n-m)\varphi]\Big), \\
    D_{\mathbf{n}\mathbf{m}} = -(V^{eg}_{\mathbf{n}\mathbf{m}}-U^{eg}_{\mathbf{n}\mathbf{m}})\sin[(n-m)\varphi], \\
    C_{\mathbf{n}\mathbf{m}} = V^{ee}_{\mathbf{n}\mathbf{m}}-V^{gg}_{\mathbf{n}\mathbf{m}},
    \end{gathered}
\end{equation}
where $\varphi=\pi\lambda_L/\lambda_c$ is the clock laser phase difference between nearest-neighbor sites ($\lambda_L$ is the lattice wavelength, $\lambda_c$ is the clock laser wavelength), $U^{\alpha\beta}_{\mathbf{n}\mathbf{m}}$ and $V^{\alpha\beta}_{\mathbf{n}\mathbf{m}}$ are $s$-wave and $p$-wave interaction parameters respectively ($\alpha,\beta=\{g,e\}$),
\begin{equation}
    U^{\alpha\beta}_{\mathbf{n}\mathbf{m}} = \frac{4\pi\hbar a_{\alpha\beta}}{M}s_{m_xm_x}s_{n_ym_y}\eta_{nm}, \quad V^{\alpha\beta}_{\mathbf{n}\mathbf{m}} = \frac{3\pi\hbar b_{\alpha\beta}^3}{M}(p_{m_xm_x}s_{n_ym_y}+s_{m_xm_x}p_{n_ym_y})\eta_{nm}.
\end{equation}
Here, $M$ is the mass of $^{87}$Sr atoms, $a_{\alpha\beta}$ and $b^3_{\alpha\beta}$ are $s$-wave scattering lengths and $p$-wave scattering volumes with $a_{eg}\equiv (a_{eg}^{+}+a_{eg}^{-})/2$ and $\beta_{eg}^{3}\equiv [(\beta_{eg}^{+})^{3}+(\beta_{eg}^{-})^{3}]/2$. $s_{nm}$, $p_{nm}$ and $\eta_{nm}$ are overlap integrals of wave functions, $s_{nm}=\int dX[\phi_n(X)]^2[\phi_m(X)]^2$, $p_{nm}=\int dX[(\partial_X\phi_n(X))\phi_m(X)-\phi_n(X)(\partial_X\phi_m(X))]^2$, $\eta_{nm}=\int dZ [W_n(Z)]^2[W_m(Z)]^2$, where $\phi_n(X)$ are harmonic oscillator wave functions, and $W_n(Z)$ are Wannier-Stark wave functions.

$\hat{H}_{\rm spectator}$ describes the interactions between clock atoms ($|e\rangle$ or $|g\rangle$ states) and spectator atoms (not in clock states).
Both the lattice-induced decay and spontaneous emission, introduce atom decay from the $|e\rangle$ and $|g\rangle$ states during clock interrogation. After a  decay event atoms can  populate other nuclear spins in the ground state manifold. This leads to non-negligible on-site $s$-wave density shift. Here we use $|a\rangle$ to label the extra ground state level (not in state $|g\rangle$) with the same nuclear spin as excited state $|e\rangle$. The interactions between $|a\rangle$ and $|e\rangle$ ($|g\rangle$) states can be described by $s$-wave scattering length $a^{-}_{eg}$ ($a_{gg}$). 
We use $|b\rangle$ to label the extra ground state levels (not in state $|g\rangle$) with different nuclear spins compare to excited state $|e\rangle$. Notice that $b$ might refer to several ground state levels, and there is no need to distinguish them because the interaction would be the same.
The interactions between $|b\rangle$ and $|e\rangle$ ($|g\rangle$) states can be described by $s$-wave scattering length $a_{eg}$ ($a_{gg}$). 
We use operators $\hat{N}^a_{\mathbf{n}}$ and $\hat{N}^b_{\mathbf{n}}$ to describe the population in $|a\rangle$ and $|b\rangle$ state for mode $\mathbf{n}$ respectively.
The Hamiltonian parameters in $\hat{H}_{\rm spectator}$ are given by
\begin{equation}
    K^a_{\mathbf{n}\mathbf{m}} = U^{eg-}_{\mathbf{n}\mathbf{m}} - U^{gg}_{\mathbf{n}\mathbf{m}}, \quad K^b_{\mathbf{n}\mathbf{m}} = U^{eg}_{\mathbf{n}\mathbf{m}} - U^{gg}_{\mathbf{n}\mathbf{m}}.
\end{equation}

Apart from Hamiltonian dynamics, we include all the dissipative terms discussed in Fig.~1 of the main text. We define fermionic annihilation operator $\hat{c}_{\mathbf{n}\alpha}$ for mode $\mathbf{n}$ and internal state $\alpha$. The Lindblad jump operators are listed below:
\begin{itemize}
    \item Spontaneous emission and lattice-induced decay
    \begin{equation}
        \sqrt{\Gamma_{eg}}\hat{c}^{\dag}_{\mathbf{n}g}\hat{c}_{\mathbf{n}e}, \quad \sqrt{\Gamma_{ea}}\hat{c}^{\dag}_{\mathbf{n}a}\hat{c}_{\mathbf{n}e}, \quad \sqrt{\Gamma_{eb}}\hat{c}^{\dag}_{\mathbf{n}b}\hat{c}_{\mathbf{n}e}, \quad \forall \mathbf{n}
    \end{equation}
    Here we assume the decay processes do not change the motional state of the atoms. This approximation is valid if the heating effect is not significant. $\Gamma_{eg}$, $\Gamma_{ea}$, $\Gamma_{eb}$ can be calculated from $\Gamma_L(0)$ and $\partial_U\Gamma_L$ in the Table I of the main text by multiplying branching ratio to different nuclear spin states.

    \item Single-body loss
    \begin{equation}
        \sqrt{\Gamma_{e}}\hat{c}_{\mathbf{n}e}, \quad \sqrt{\Gamma_{g}}\hat{c}_{\mathbf{n}g}, \quad \sqrt{\Gamma_{g}}\hat{c}_{\mathbf{n}a}, \quad \sqrt{\Gamma_{g}}\hat{c}_{\mathbf{n}b}, \quad \forall \mathbf{n}
    \end{equation}
    $\Gamma_e$ and $\Gamma_g$ are taken from the Table I of the main text. 

    \item Two-body ee loss
    \begin{equation}
        \sqrt{\Gamma_{\mathbf{n}\mathbf{m}}}\hat{c}_{\mathbf{n}e}\hat{c}_{\mathbf{m}e}, \quad \forall \mathbf{n},\mathbf{m},
    \end{equation}
    where ($\beta_{ee}^3$ is the inelastic $p$-wave scattering volume)
    \begin{equation}
        \Gamma_{\mathbf{n}\mathbf{m}}=\frac{3\pi\hbar^2\beta_{ee}^3}{2M}(p_{m_xm_x}s_{n_ym_y}+s_{m_xm_x}p_{n_ym_y})\eta_{nm}.
    \end{equation}
    The thermal average of $\Gamma_{\mathbf{n}\mathbf{m}}$ agrees with $\tilde{\Gamma}_{ee}\kappa$ in Table I of the main text, where $\kappa$ is a prefactor determined by density and temperature of the cloud.
    
\end{itemize}

In the calculation procedure described above, we are simply treating the lattice-induced scattering to $^{3}P_2$ states as atom loss. In experiment, atoms can also stay in the long-lived $^{3}P_2$ states and increase the strength of spectator interaction. However, the scattering lengths of the $^{3}P_2$ states are unknown. To take account of this effect, we consider a rescaling of spectator interaction $\hat{H}_{\rm spectator}\rightarrow \xi\hat{H}_{\rm spectator}$, with $\xi$ a fitting parameter. We choose $\xi=3$ in comparison with experimental data.

\begin{figure}[h!]
    \includegraphics{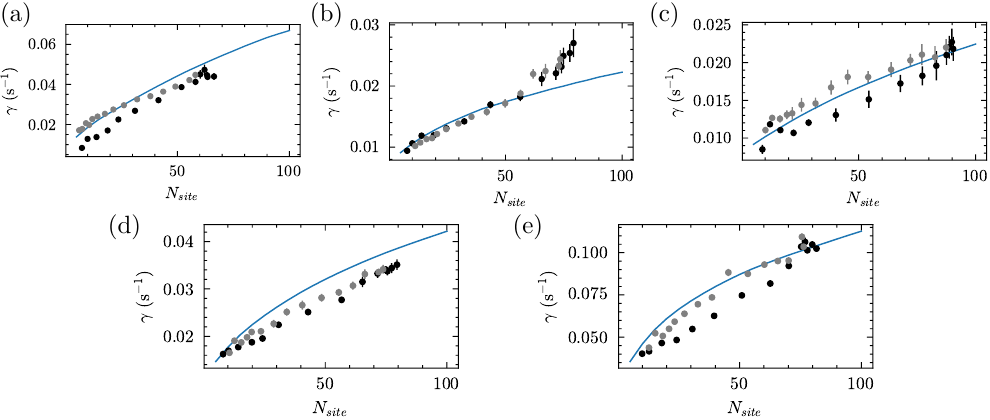}
    \caption{\label{fig:supp_depth_nonlinear} $N_{site}$ dependence of the contrast decay rate $\gamma$ for at (a) 5\Er{}, (b) 8\Er{}, (c) 11\Er{}, (d) 23\Er{}, and (e) 69\Er{}. The black markers are data points represent the regions with higher initial spin purity, and the gray markers show the other part of the sample. The error bars show 68\% confidence interval. The blue solid lines are DDTWA simulation.
    }
\end{figure}

In Fig.~\ref{fig:supp_depth_nonlinear}, we show the comparison between DDTWA simulations and experimental data for contrast decay rate $\gamma$. 
Both DDTWA simulations and experimental data show a clear dependence on the atom number per site ($N_{site}$) and exhibit a slight nonlinearity. 
To account for this nonlinearity in our linear extrapolation in the main text, we perform fits over two different ranges of $N_{\text{site}}$ and include the difference between these fits as an additional source of uncertainty.
We find a spatial inhomogeneity on observed contrast decay rate, presumably due to the small spatial inhomogeneity of the initial spin impurity. This could bias the analysis and complicates the modeling. Therefore, we use the half of the region with better spin impurity for the analysis in the main text.

\section{Lower bound of sensitivity for ellipse fitting}
In ellipse fitting, we measure the excitation fraction $x_i,y_i$ in two separate regions  ($i=1,\cdots, n$ is labeling independent measurements), and use this information to determine the differential phase $\phi$. The expectation value of $x_i,y_i$ is given by
\begin{equation}\label{eq:ellipse}
    \mathrm{E}(x_i) = \frac{C}{2}\cos(\theta_i-\phi/2)+D, \quad \mathrm{E}(y_i) = \frac{C}{2}\cos(\theta_i+\phi/2)+D.
\end{equation}
In particular, we focus on the case where the common phase $\theta_i$ is randomly distributed between $0$ and $2\pi$ due to laser phase noise, so the values $(x_i,y_i)$ are forming an ellipse. Here, $C$ is the Ramsey contrast, and $D$ is the center of the ellipse. 
The conditional probability for the repeated measurements is given by 
\begin{equation}
    p(\mathbf{x},\mathbf{y}|\phi,\bm{\theta}) = \prod_i p(x_i,y_i|\phi,\theta_i),
\end{equation}
where $p(x_i,y_i|\phi,\theta_i)$ is the probability of measurement outcome $x_i,y_i$ conditioned on parameters $\phi$ and $\theta_i$. In principle, $x_i,y_i$ are also conditioned on parameters $C$ and $D$, here we omit them to shorten the notations.
For simplicity, we assume the quantum state is separable between the two regions, and that in each region can be approximated by a Gaussian state, leading to
\begin{equation}
    \begin{aligned}
    p(x_i,y_i|\phi,\theta_i)&=p(x_i|\phi,\theta_i)p(y_i|\phi,\theta_i)\\
    &\approx \frac{1}{\sqrt{2\pi \mathrm{Var}(x_i)}}\exp\bigg(-\frac{1}{2}\frac{(x_i-\mathrm{E}(x_i))^2}{\mathrm{Var}(x_i)}\bigg) \times \frac{1}{\sqrt{2\pi \mathrm{Var}(y_i)}}\exp\bigg(-\frac{1}{2}\frac{(y_i-\mathrm{E}(y_i))^2}{\mathrm{Var}(y_i)}\bigg),
    \end{aligned}
\end{equation}
in which $\mathrm{Var}(x_i)$ depends on $\theta_i-\phi/2$, and $\mathrm{Var}(y_i)$ depends on $\theta_i+\phi/2$. 

The lower bound of sensitivity for differential phase $\phi$ can be estimated by the classical Cramér–Rao bound. A straightforward way to apply the classical Cramér–Rao bound is to integrate out the common phases $\theta_i$ \cite{corgierSqueezingenhancedAccurateDifferential2025},
\begin{equation}
    p(\mathbf{x},\mathbf{y}|\phi) = \prod_i p(x_i,y_i|\phi), \quad  p(x_i,y_i|\phi) = \int_0^{2\pi} \frac{d\theta_i}{2\pi} p(x_i,y_i|\phi,\theta_i),
\end{equation}
and then construct the classical Fisher information,
\begin{equation}
    F_{\phi}=\int d\mathbf{x}d\mathbf{y} \; p(\mathbf{x},\mathbf{y}|\phi) \bigg(\frac{\partial \ln p(\mathbf{x},\mathbf{y}|\phi)}{\partial \phi}\bigg)^2 = n \int dxdy \; p(x,y|\phi) \bigg(\frac{\partial \ln p(x,y|\phi)}{\partial \phi}\bigg)^2,
\end{equation}
in which the last step in based on our assumption of independent measurements. The classical Cramér–Rao bound leads to
\begin{equation}
    \mathrm{Var}(\phi) \geq F_{\phi}^{-1}.
    \label{eq:lowerbound1}
\end{equation}
This approach provides the tightest lower bound for $\mathrm{Var}(\phi)$, while analytical calculation for this bound is hard to achieve.

Here we would like to provide a slightly loose lower bound for $\mathrm{Var}(\phi)$, where we can have more analytical insights. We start from the classical Fisher information matrix for all the parameters $\phi,\bm{\theta}$,
\begin{equation}
    F = \begin{pmatrix}
        F_{\phi\phi} & F_{\phi\theta_1} & \cdots & F_{\phi\theta_n}\\
        F_{\theta_1\phi} & F_{\theta_1\theta_1} & \cdots & F_{\theta_1\theta_n}\\
        \vdots & \vdots & \ddots & \vdots\\
        F_{\theta_n\phi} & F_{\theta_n\theta_1} & \cdots & F_{\theta_n\theta_n}\\
    \end{pmatrix},
\end{equation}
where
\begin{equation}
    \begin{aligned}
    F_{\phi\phi} &= \int d\mathbf{x}d\mathbf{y} \; p(\mathbf{x},\mathbf{y}|\phi,\bm{\theta}) \bigg(\frac{\partial \ln p(\mathbf{x},\mathbf{y}|\phi,\bm{\theta})}{\partial \phi}\bigg)^2 \\
    &\approx \sum_i\bigg(\frac{(\partial_{\phi}\mathrm{E}(x_i))^2}{\mathrm{Var}(x_i)}+\frac{(\partial_{\phi}\mathrm{E}(y_i))^2}{\mathrm{Var}(y_i)}\bigg).
    \end{aligned}
\end{equation}
\begin{equation}
    \begin{aligned}
    F_{\phi\theta_i} = F_{\theta_i\phi} &= \int d\mathbf{x}d\mathbf{y} \; p(\mathbf{x},\mathbf{y}|\phi,\bm{\theta}) \bigg(\frac{\partial \ln p(\mathbf{x},\mathbf{y}|\phi,\bm{\theta})}{\partial \phi}\bigg)\bigg(\frac{\partial \ln p(\mathbf{x},\mathbf{y}|\phi,\bm{\theta})}{\partial \theta_i}\bigg)\\
    &\approx \frac{(\partial_{\phi}\mathrm{E}(x_i))(\partial_{\theta}\mathrm{E}(x_i))}{\mathrm{Var}(x_i)}+\frac{(\partial_{\phi}\mathrm{E}(y_i))(\partial_{\theta}\mathrm{E}(y_i))}{\mathrm{Var}(y_i)},
    \end{aligned}
\end{equation}
\begin{equation}
    \begin{aligned}
    F_{\theta_i\theta_j} &= \int d\mathbf{x}d\mathbf{y} \; p(\mathbf{x},\mathbf{y}|\phi,\bm{\theta}) \bigg(\frac{\partial \ln p(\mathbf{x},\mathbf{y}|\phi,\bm{\theta})}{\partial \theta_i}\bigg)\bigg(\frac{\partial \ln p(\mathbf{x},\mathbf{y}|\phi,\bm{\theta})}{\partial \theta_j}\bigg)\\
    &\approx \bigg(\frac{(\partial_{\theta_i}\mathrm{E}(x_i))^2}{\mathrm{Var}(x_i)}+\frac{(\partial_{\theta_i}\mathrm{E}(y_i))^2}{\mathrm{Var}(y_i)}\bigg)\delta_{ij}.
    \end{aligned}
\end{equation}
The calculation above is based on the nature of Gaussian state and the large atom number limit in each region, so we can drop the derivatives for $\mathrm{Var}(x_i)$ and $\mathrm{Var}(y_i)$.

Based on the multivariate classical Cramér–Rao bound and the fact that the diagonal element of a positive semi-definite matrix is positive semi-definite, we have
\begin{equation}
    \mathrm{Var}(\phi)\geq (F^{-1})_{\phi\phi} = \bigg(F_{\phi\phi}-\sum_i F_{\phi\theta_i}F_{\theta_i\theta_i}^{-1}F_{\theta_i\phi}\bigg)^{-1}.
\end{equation}
Here the last step is based on the inverse of a block matrix. Notice that $\partial_{\theta_i}E(x_i) = -2\partial_{\phi}E(x_i)$, $\partial_{\theta_i}E(y_i) = 2\partial_{\phi}E(y_i)$,
we have
\begin{equation}
    \begin{aligned}
    F_{\phi\phi}-\sum_i F_{\phi\theta_i}F_{\theta_i\theta_i}^{-1}F_{\theta_i\phi} &\approx  4\sum_i\bigg(\frac{(\partial_{\phi}\mathrm{E}(x_i))^2}{\mathrm{Var}(x_i)}+\frac{(\partial_{\phi}\mathrm{E}(y_i))^2}{\mathrm{Var}(y_i)}\bigg)^{-1}\frac{(\partial_{\phi}\mathrm{E}(x_i))^2}{\mathrm{Var}(x_i)}\frac{(\partial_{\phi}\mathrm{E}(y_i))^2}{\mathrm{Var}(y_i)}\\
    &=\frac{C^2}{4}\sum_i \frac{1}{\csc^2(\theta_i-\phi/2)\mathrm{Var}(x_i)+\csc^2(\theta_i+\phi/2)\mathrm{Var}(y_i)}\\
    &=\frac{nC^2}{4}\int_0^{2\pi} \frac{d\theta}{2\pi}\frac{1}{\csc^2(\theta-\phi/2)\mathrm{Var}(x)+\csc^2(\theta+\phi/2)\mathrm{Var}(y)},
    \end{aligned}
\end{equation}
in which $\mathrm{Var}(x)$ and $\mathrm{Var}(y)$ are functions of $\theta$ and $\phi$. So we have 
\begin{equation}
    \mathrm{Var}(\phi)\geq \frac{4}{nC^2} \bigg[\int_0^{2\pi} \frac{d\theta}{2\pi}\bigg(\frac{1}{\csc^2(\theta-\phi/2)\mathrm{Var}(x)+\csc^2(\theta+\phi/2)\mathrm{Var}(y)}\bigg)\bigg]^{-1}.
     \label{eq:lowerbound2}
\end{equation}
Note that this formula agrees with the formula listed in the SOM of Ref.~\cite{martiImagingOpticalFrequencies2018}.

\begin{figure}[th!]
    \includegraphics[width=16cm]{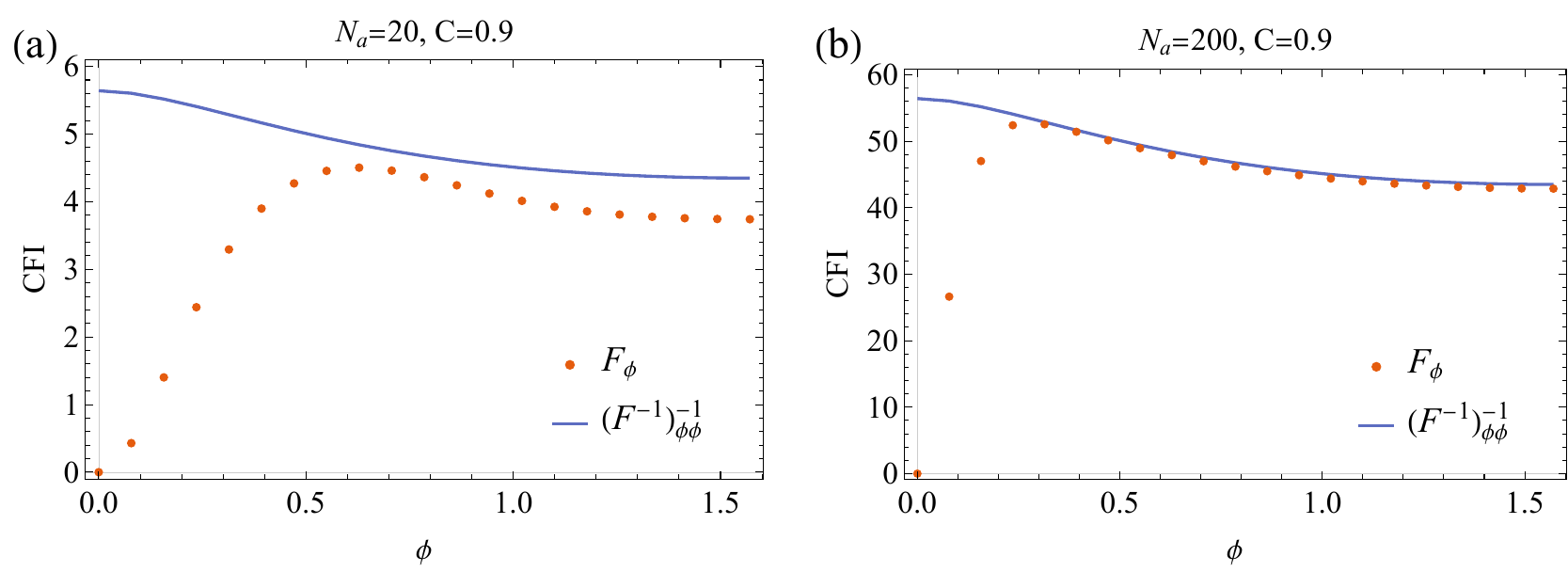}
    \caption{\label{fig:cfi} Comparison Eq.~(\ref{eq:lowerbound1}) and Eq.~(\ref{eq:lowerbound2}) for differential phase $\phi$ with (a) $N_a=20$ and (b) $N_a=200$. We consider the quantum state is a product state with contrast $C=0.9$.
    }
\end{figure}

Now we would like provide a benchmarking between the two different lower bounds (see Eq.~(\ref{eq:lowerbound1}) and Eq.~(\ref{eq:lowerbound2})). We consider the quantum state is a product state with contrast $C=0.9$, so the variances for excitation fraction $x$ and $y$ are given by $\mathrm{Var}(x) = \mathrm{E}(x)(1-\mathrm{E}(x))/N_a$, $\mathrm{Var}(y) = \mathrm{E}(y)(1-\mathrm{E}(y))/N_a$, where $N_a$ is the atom number for each region. In Fig.~\ref{fig:cfi}, we compare Eq.~(\ref{eq:lowerbound1}) and Eq.~(\ref{eq:lowerbound2}) for $N_a=20$ and $N_a=200$. We find that Eq.~(\ref{eq:lowerbound1}) and Eq.~(\ref{eq:lowerbound2}) agrees for most of the $\phi$ except a small region near $\phi=0$, and the region of disagreement is shrinking as we increase $N_a$. Considering $N_a\sim 10^5$ in our experiment, we believe that Eq.~(\ref{eq:lowerbound2}) is a good approximation for Eq.~(\ref{eq:lowerbound1}) unless getting very close to $\phi=0$.

A physical interpretation for $F_{\phi}\sim 0$ near $\phi=0$ can be gained by noticing that the ellipse is becoming a diagonal line, and the quantum noise makes the data points to be either above or below that line. In that case  $\phi>0$ and $\phi<0$  cases cannot be distinguished well and therefore we are not extracting any information using this method. 
Since the quantum noise will decrease as we increase $N_a$, then  the value of  $\phi$ for which  we can  distinguish   if  the data points are above or below the diagonal line  decreases with increasing $N_a$.

\section{Ellipse fitting method}
Extracting the differential phase $\phi$ from the ellipse fitting can be done in various ways~\cite{corgierSqueezingenhancedAccurateDifferential2025,stocktonBayesianEstimationDifferential2007,fosterMethodPhaseExtraction2002,
collettEllipseFittingInterferometry2015}. We employ two different methods: (1) least squares with geometric distance (LSGD) and (2) Bayesian inference (BI). For estimating uncertainty, we use the jackknife method~\cite{martiImagingOpticalFrequencies2018} for both LSGD and BI. BI with Markov chain Monte Carlo (MCMC) provides an approximate posterior distribution of $\phi$~\cite{abril-plaPyMCModernComprehensive2023}. We use the standard deviation of the marginalized posterior distribution as the uncertainty and compare it with jackknife method.

LSGD use a cost function based on sum of geometric distances
\begin{equation}
    \mathrm{Cost} = \sum_{i=1}^n d_i = \sum_{i=1}^n \left(x_i - E(x_i) \right)^2 + \left(y_i - E(y_i) \right)^2,
\end{equation}
where $d_i$ is the geometric distance between the data point $(x_i,y_i)$ with index $i=1, 2, ..., n$ and the points on an ellipse, and fitted values $(E(x_i) ,E(y_i) )$ are from Eqn.~\eqref{eq:ellipse} for a given $n+4$ parameters $\left\{C,\phi, \bm{D}, \bm{\theta}\right\}$ ($D$ is separated for $x, y$).

For BI, we model the data points $\{(x_i,y_i)\}$ as
\begin{equation}
    (x_i,y_i)|C, \phi, \bm{D}, \theta_i, N \sim \left(\mathcal{N}(E(x_i),\sigma_{x_i}^2(N_a)), \mathcal{N}(E(y_i),\sigma_{y_i}^2(N_a))\right), 
\end{equation}
where $\mathcal{N}(\mu,\sigma^2)$ is the normal distribution with mean $\mu$ and variance $\sigma^2$. $\sigma_{x_i}^2(N_a) = x_i(1-x_i)/N_a$ is a variance estimation assuming large number limit of a coherent spin state, and we also infer an effective atom number $N_a$ per ensemble from the data. This approximation is only valid for a sufficiently large atom number~\cite{ecknerRealizingSpinSqueezing2023}. For the prior distribution, we first perform a LSGD fit to the data, and use the fitted parameters as the mean values of the normally distributed prior. The prior sensitivity is checked by varying the prior width for a given data set. The posterior distribution is insensitive for the width change from 0.05 to 0.0001 for $C$, $\phi$, and $\bm{\theta}$ for the data set used in the main text and the simulated data presenting here. 

Estimating the uncertainty of $\phi$ using the jackknife method is done by removing one data point at index $i$ and performing the LSGD or BI to the remaining data points~\cite{martiImagingOpticalFrequencies2018}. The bias corrected estimator for $\phi$ for each $i$ is given by
\begin{equation} \label{eq:jackknife}
    \phi_i^{\mathrm{JK}}=n \bar{\phi}^{\mathrm{JK}}-(n-1) \bar{\phi}_{\neq i}^{\mathrm{JK}},
\end{equation}
where $\bar{\phi}^{\mathrm{JK}}$ is the $\phi$ obatined from using all data points, and $\bar{\phi}_{\neq i}^{\mathrm{JK}}$ is obtained from using all data points except $i$.

We check the performance of the fitting methods with simulated data assuming a coherent spin state and compare with the result from the experimental data.
\begin{figure}[th!]
    \includegraphics[]{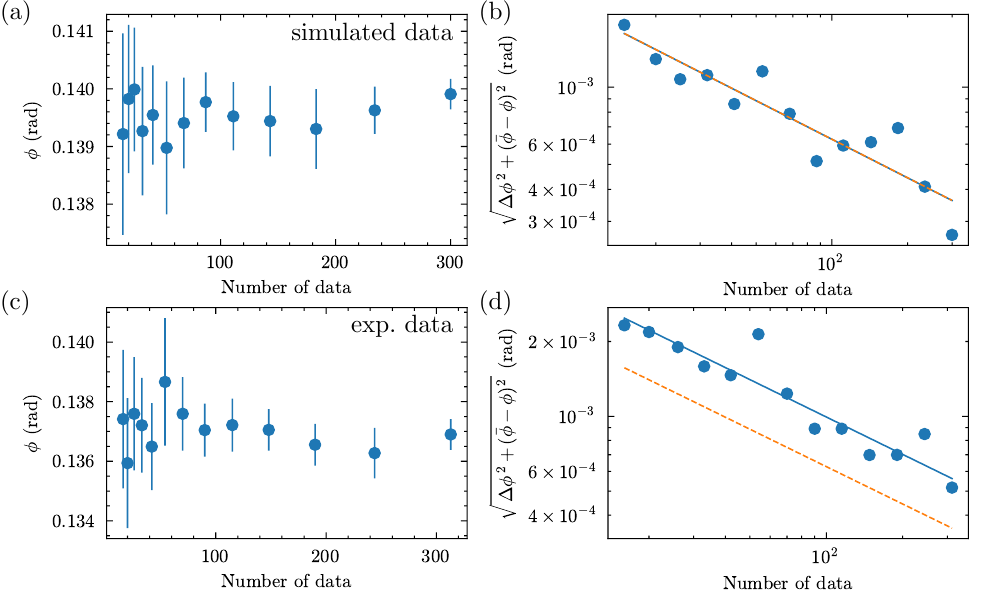}
    \caption{\label{fig:supp_timebinadev}
    Bayesian estimation of $\phi$ with increasing size of the data set from product state simulation (a, b) and from experimental data (c, d). (a) Estimated $\phi$ as a function of the number of data points. The marker represents the mean value of the posterior distribution, and the error bar represents the 68\% confidence interval. (b) Uncertainty of the $\phi$ as a function of the number of data points. The blue solid line is a fit ($\propto 1/\sqrt{n}$) to the data and the orange dashed line is the theoretical estimation (6.3~mrad for 1 data). $\bar{\phi}$ is $\phi$ with full data set, used for estimating the bias. (c, d) are the same as (a, b) but with experimental data. Experimental data shows 1.6 times larger noise (the blue solid line).
    }
\end{figure}
Figure~\ref{fig:supp_timebinadev} shows the results of the BI for $\phi$ with increasing size of the data set. We simulate data assuming $\{C=0.74, \phi=0.14, N=160,000, D=0.5\}$, close to the experimental condition, and $\theta_i$ is randomly sampled from a uniform distribution between $0$ and $2\pi$. For the theoretical estimation, we use Eqn.~\eqref{eq:lowerbound2}. For the simulated data, we find that the stability obtained from the posterior distribution of $\phi$ roughly matches to the theoretical prediction. The same method applied to the experimental data shows a larger uncertainty. The scatters on the variance is likely from the finite sample size. 
\begin{figure}[th!]
    \includegraphics[]{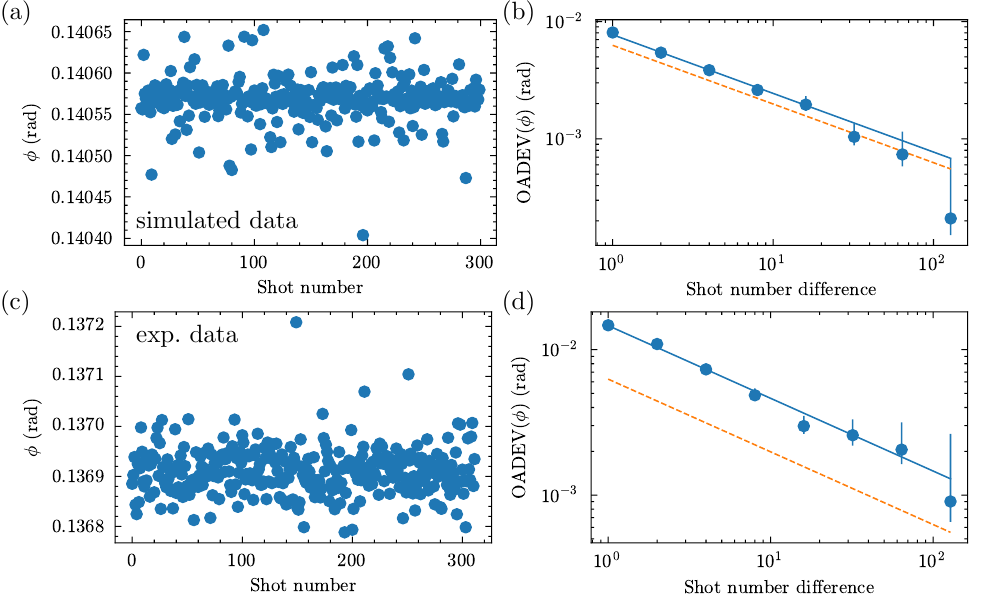}
    \caption{\label{fig:supp_jackknife}
    BI with jackknife method with data set from product state simulation (a, b) and from experimental data (c, d). (a) Estimated $\phi$ from Eqn.~\eqref{eq:jackknife}. (b) Allan deviation of $\phi$ The blue solid line is a fit to the data and the orange dashed line is the theoretical estimation (6.3~mrad for 1 data). (c, d) are the same as (a, b) but with experimental data. Experimental data shows 2.4 times larger noise (the blue solid line).
    }
\end{figure}

In Fig.~\ref{fig:supp_jackknife}, we show the jackknife estimation with BI. For simulated data set, the jackknife method gives a similar result to the theoretical estimation. 

\begin{figure}[th!]
    \includegraphics[]{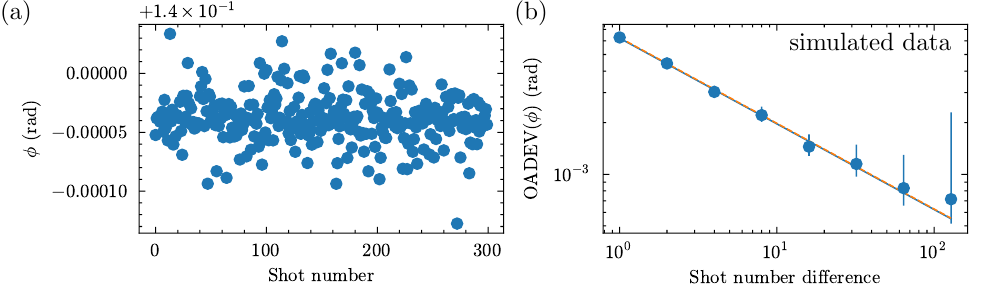}
    \caption{\label{fig:supp_geo_jackknife}
    Jackknife estimation with LSGD for a product state simulation. (a) Estimated $\phi$ from Eqn.~\eqref{eq:jackknife}. (b) Allan deviation of $\phi$ The blue solid line is a fit to the data and the orange dashed line is the theoretical estimation (6.3~mrad for 1 data). The result for experimental data is shown in the main text.
    }
\end{figure}

Finally, we show the jackknife estimation with LSGD in Fig.~\ref{fig:supp_geo_jackknife}. The result is similar to the BI method. The result with the experimental data is shown in the main text as Fig.~3(e).
We generally find a good agreement between different analysis methods, with the remaining discrepancies likely arising from the finite sample size of $n\sim300$. The DDTWA simulation indicates a higher noise level compared to simulations based on coherent spin states, which is primarily due to the stochastic generation of spectator atoms (statistical mixture). It is thus challenging to develop a noise model for Bayesian analysis using DDTWA simulated data. Consequently, we decided to use the most straightforward method, LSGD with jackknife, for the analysis presented in the main text.

\clearpage
\bibliography{main}